\documentclass{PoS}
\usepackage{subfigure}
\usepackage{graphicx,picture,calc}

\title{Matrix Product States for Lattice Field Theories}

\ShortTitle{Matrix Product States for Lattice Field Theories}

\author{\speaker{M. C.~Ba\~nuls}$^a$, K.~Cichy$^{bc}$, J. I.~Cirac$^a$, K.~Jansen$^b$ and H.~Saito$^{bd}$
\\
\llap{$^a$}Max-Planck-Institut f\"ur Quantenoptik,
 Hans-Kopfermann-Str. 1, 85748 Garching, Germany\\
\llap{$^b$}NIC, DESY Zeuthen, Platanenallee 6, 15738 Zeuthen, Germany\\
\llap{$^c$}Adam Mickiewicz University, Faculty
of Physics, Umultowska 85, 61-614 Poznan, Poland\\
\llap{$^d$}Graduate School of Pure and Applied Sciences, University of Tsukuba, Tsukuba, Ibaraki 305-8571, Japan\\
E-mail: \email{banulsm@mpq.mpg.de}, \email{krzysztof.cichy@desy.de}, \email{ignacio.cirac@mpq.mpg.de},
\email{karl.jansen@desy.de}, \email{hana.saito@desy.de}
}

\abstract{The term Tensor Network States (TNS) refers to a number of families of states that represent
different
ans\"atze for the efficient description of the state of a quantum many-body system. Matrix Product
States
(MPS) are one particular case of TNS, and have become the most precise tool for the numerical study of
one
dimensional quantum many-body systems, as the basis of the Density Matrix Renormalization Group method.
Lattice Gauge Theories (LGT), in their Hamiltonian version, offer a challenging scenario for these techniques.
While the dimensions and sizes of the systems amenable to TNS studies are still far from those achievable
by 4-dimensional 
LGT tools,
 Tensor Networks can be readily used for problems which more standard techniques, such as Markov chain
Monte Carlo simulations, cannot easily tackle. Examples of such problems are the presence of a chemical
potential or out-of-equilibrium dynamics. We
have explored the performance of Matrix Product States in the case of the Schwinger model, as a
widely used testbench for lattice techniques. Using finite-size, open boundary MPS, we are able to
determine the low energy states of the model in a fully non-perturbative manner. The precision achieved
by
the method allows for accurate finite size and continuum limit extrapolations of the ground state energy,
but also of the chiral condensate
and the mass gaps, thus showing the feasibility of these techniques for gauge theory problems.}

\FullConference{31st International Symposium on Lattice Field Theory - LATTICE 2013\\
		July 29 - August 3, 2013\\
		Mainz, Germany}

\begin{document}

\section{Introduction}


Tensor Network States (TNS) are ans{\"a}tze for the description of 
quantum many-body systems, whose common characteristic is the efficient encoding
of the entanglement pattern present in the physical states.
The Matrix Product State (MPS) ansatz~\cite{aklt88,fannes92fcs,vidal03eff,verstraete04dmrg} 
explains  the enormous success 
of the Density Matrix Renormalization Group (DMRG) algorithm~\cite{white92dmrg,schollwoeck11age}
at computing ground states of quantum spin chains.
The insight gained from quantum information theory~\cite{cirac09rg}
 allowed the extension of tensor network (TN) methods to 
 higher dimensions and  
 dynamical problems. 
 In the last years, TN algorithms are hence seen as a very promising approach for strongly 
 correlated quantum many-body problems, with the potential to attack questions 
which are hard for other techniques, such as Markov chain Monte Carlo simulations, in particular,
fermionic or frustrated spin systems, and 
out-of-equilibrium dynamics.
 
 Quantum field theories, in particular in their lattice formulation, 
 open a particularly interesting realm of application
 for these non-perturbative techniques.
 Lattice Gauge Theory (LGT), despite its enormous  
 theoretical and technical development, 
 has limited 
 applicability when 
 dealing with dynamics or finite density,
 and TN methods could in the future  represent an alternative approach, capable to
 overcome some of these problems.
Although several TNS approaches have already been tried on 
different lattice field theory problems~\cite{byrnes02dmrg,sugihara05gaugeinv,tagliacozzo11,banuls13schwinger},
there is not yet a systematic exploration of the power of these techniques to tackle the questions that 
standard LGT methods face.

In~\cite{banuls13schwinger} we showed the suitability of the MPS ansatz to describe not only the ground
state, but also the
excitations of the lattice Schwinger model, and to produce continuum limit extrapolations
of the mass spectrum whose precision can compete with that of other numerical techniques.
Despite the very precise estimation of the energy, other observables~\cite{legeza96accuracy}
might be more sensitive to truncation errors and yield a considerably worse estimation.
Here, we complement the study in~\cite{banuls13schwinger} with the explicit calculation of the chiral
condensate,
and show that the MPS method also provides
accurate continuum limit extrapolations,
beyond the precision reached by perturbative
~\cite{adam97,hosotani98} or non-perturbative methods~\cite{deForcrand98,duerr05scaling}.

\section{The model}

To probe the suitability of MPS methods for LGT problems,  the lattice 
Schwinger model~\cite{schwinger62,hamer82} in its 
Kogut-Susskind Hamiltonian formulation~\cite{kogut75sc} is used as  test bench. 
Via a Jordan-Wigner transformation, the fermionic degrees of freedom can be mapped to spin variables~\cite{Banks:1975gq}.
The additional gauge degrees of freedom sitting on the links are constrained 
by Gauss' law, 
which in the case of open boundary conditions (OBC) considered here
completely fixes the electric field up to a constant.
This is enough to fully eliminate the gauge variables and work
with the following dimensionless long-range spin model
 \cite{hamer97free}
\begin{equation}
H= x\sum_{n=0}^{N-2} \left [ \sigma_n^+\sigma_{n+1}^- +  \sigma_n^-\sigma_{n+1}^+ \right ]+\frac{\mu}{2}\sum_{n=0}^{N-1} \left [ 1 + (-1)^n \sigma_n^z \right ]
+\sum_{n=0}^{N-2} \left [ \ell +\frac{1}{2}\sum_{k=0}^n ((-1)^k+\sigma_k^{z})\right ]^2,
\label{eq:H}
\end{equation}
where the Hamiltonian parameters $x=\frac{1}{g^2 a^2}$ and $\mu=\frac{2 m}{g^2 a}$ are 
expressed in terms of the lattice spacing, $a$, the coupling, $g$, and fermion mass $m$.
The parameter $\ell$ represents the boundary electric field on the link to the left of site $0$, which can describe the background field.
For OBC  $\ell$ is non-dynamical, and in the following we choose it to be zero.  
Therefore, we can use a tensor product spin basis $|i_0\ldots i_{N-1}\rangle$, with $i_m=\{0,1\}$, 
to describe the states of the system.

\section{Chiral condensate}

All observables can be written in terms of spin operators. In particular, the fermionic condensate,
$\langle \overline{\Psi}(x) \Psi(x) \rangle/g$ reduces to
the ground state expectation value of $\hat{\Sigma} = \frac{\sqrt{x}}{N} \sum_n
(-1)^n \frac{1+\sigma_n^z}{2}$.
The naively computed chiral condensate, $\Sigma=\langle \mathrm{GS} | \hat{\Sigma} | \mathrm{GS} \rangle$
diverges logarithmically in the limit $a\to 0$ for non-vanishing fermion
mass~\cite{deForcrand98,duerr05scaling,Christian:2005yp}.
This divergence is already present in the free case, where the theory is exactly solvable. Indeed, 
in the non-interacting case (\ref{eq:H}) reduces to the XY spin model in a staggered magnetic field,
whose ground state energy, in the case of OBC, reads
$E_0=\frac{\mu}{2}N-\sum_{q=1}^{N/2} \sqrt{\mu^2+4 x^2 \cos^2\frac{q \pi}{N+1}}$.
The expectation value of $\hat{\Sigma}$ can then be computed from the $\frac{d E_0}{d\mu}$ as
\begin{equation}
\Sigma_{\mathrm{free}}=\frac{\sqrt{x}}{N}\sum_{q=1}^{N/2}\frac{\mu}{\sqrt{\mu^2+4 x^2 \cos^2 \frac{q \pi}{N+1}}}.
\label{eq:freecond}
\end{equation}

We may use the exact value of the free condensate to subtract the divergence from the observable computed in the interacting case. 
Our numerical procedure extracts the continuum limit by first making the physical volume infinite at constant lattice spacing 
(equivalent to the thermodynamic limit of the discrete problem) and then extrapolating to $x\to\infty$. The divergence is only 
present in the second step, when the lattice regulator vanishes, while the first limit can be evaluated via a finite 
volume extrapolation. 
In the free case we can actually extract analytically the exact value of $\Sigma_{\mathrm{free}}$ in the bulk, at fixed $(m/g,\,x)$,
when (\ref{eq:freecond}) can be evaluated as an integral, to yield
\begin{equation}
\Sigma_{\mathrm{free}}^{\mathrm{(bulk)}}(m/g,x)=\frac{m}{\pi g}\frac{1}{\sqrt{1+\frac{m^2}{g^2 x}}} \mathrm{K}\left(\frac{1}{1+\frac{m^2}{g^2x}}\right),
\label{eq:freecondbulk}
\end{equation}
where $\mathrm{K}(u)$ is the complete elliptic integral of the first kind \cite{abramowitz}.
Expanding this expression for $x\to\infty$ shows a divergent logarithmic term $\frac{1}{2\pi}\frac{m}{g}\log x$. 
The divergence  can be substracted from the interacting chiral condensate after the infinite volume extrapolation 
using the exactly computed free condensate (\ref{eq:freecondbulk}).
The interacting case may nevertheless introduce further logarithmic corrections to higher orders in
$\frac{1}{x}$ that need to be taken into account 
in the continuum limit extrapolation.

\section{Matrix Product State Methods}

We approximate the ground state of the Hamiltonian (\ref{eq:H}) 
by a  MPS, i.e. a state of the form 
$|\Psi\rangle = \sum_{i_0,\ldots i_{N-1}=0}^{d-1} \mathrm{tr}(A_0^{i_0}\ldots A_{N-1}^{i_{N-1}}) |i_0 \ldots i_{N-1}\rangle$,
where $d=2$ is the dimension of the local Hilbert space for each chain site.
The state is completely determined by the $D$-dimensional matrices $A_k^{i}$,
and hence the bond dimension, $D$, controls the number of parameters in the ansatz.
 A MPS approximation to the ground state with fixed $D$ can be found variationally  
by successively minimizing the energy with respect to each of the individual 
tensors, and iterating the procedure until convergence~\cite{verstraete04dmrg,schollwoeck11age}.
The MPS ansatz is known to provide good approximations to the ground states of local gapped Hamiltonians,
but has been used to more general models. In~\cite{banuls13schwinger} we found, indeed,
that the ansatz is appropriate for the Schwinger model, accurately reproducing 
not only the ground state energy, but also the mass spectrum.

Applying the variational MPS algorithm with open boundary conditions
we obtained an ansatz for the ground state for various sets of parameters $(m/g,x,N,D)$.
We studied four different values of the fermion mass, $m/g=0,\,0.125,\,0.25,\,0.5$,
and for each of them probed $x\in[20,\,600]$. In order to extract the bulk limit for each given $(m/g,\,x)$,
we solved the ground state problem for five different system sizes  $N\geq 20\sqrt{x}$,
to ensure large enough physical volumes. For each system size, we ran the algorithm for bond dimensions
$D\in[20,\,140]$, and stopped the iteration when the relative change in energy from one sweep to the next was below $\epsilon=10^{-12}$.

After the variational algorithm has converged for a particular set of parameters, $(m/g,\,x,\,N,\,D)$,
we obtain a MPS approximation to the ground state, from which we can easily compute expectation values
of local operators or their tensor products.
In order to extract the continuum limit properties for different fermion masses, we need to perform
successive extrapolations of the corresponding observables  in the bond dimension, $D\to \infty$, 
the system size, $N\to \infty$, and the lattice spacing $\frac{1}{\sqrt{x}}=ga\to 0$.
A detailed description of the numerical method and the corresponding error analysis can be found in~\cite{banuls13schwinger},
where we performed this procedure to extract the the ground state energy density and the mass spectrum.

\begin{figure}[floatfix]
\begin{minipage}[b]{.4\columnwidth}
\subfigure[$m/g=0$]{
\includegraphics[height=.9\columnwidth]{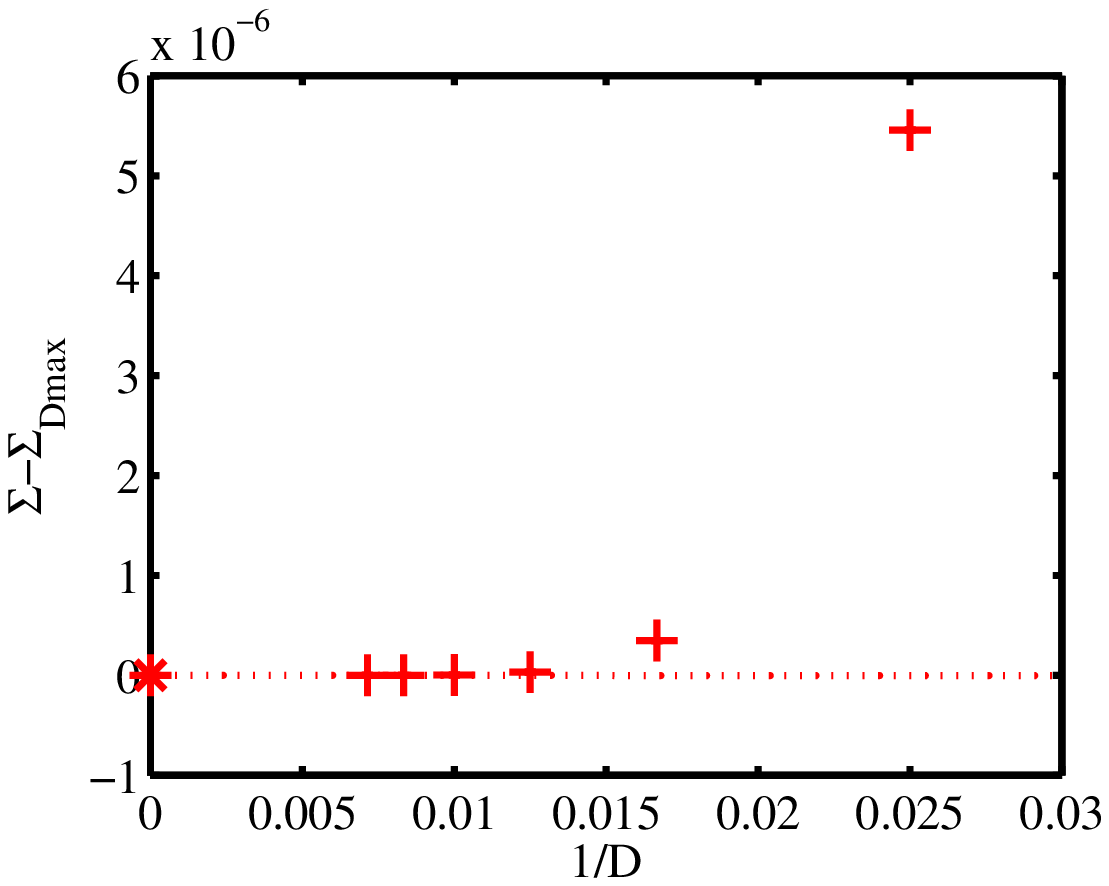}
}
\end{minipage}
\hspace{.1\columnwidth}
\begin{minipage}[b]{.4\columnwidth}
\subfigure[$m/g=0.25$]{
\includegraphics[height=.9\columnwidth]{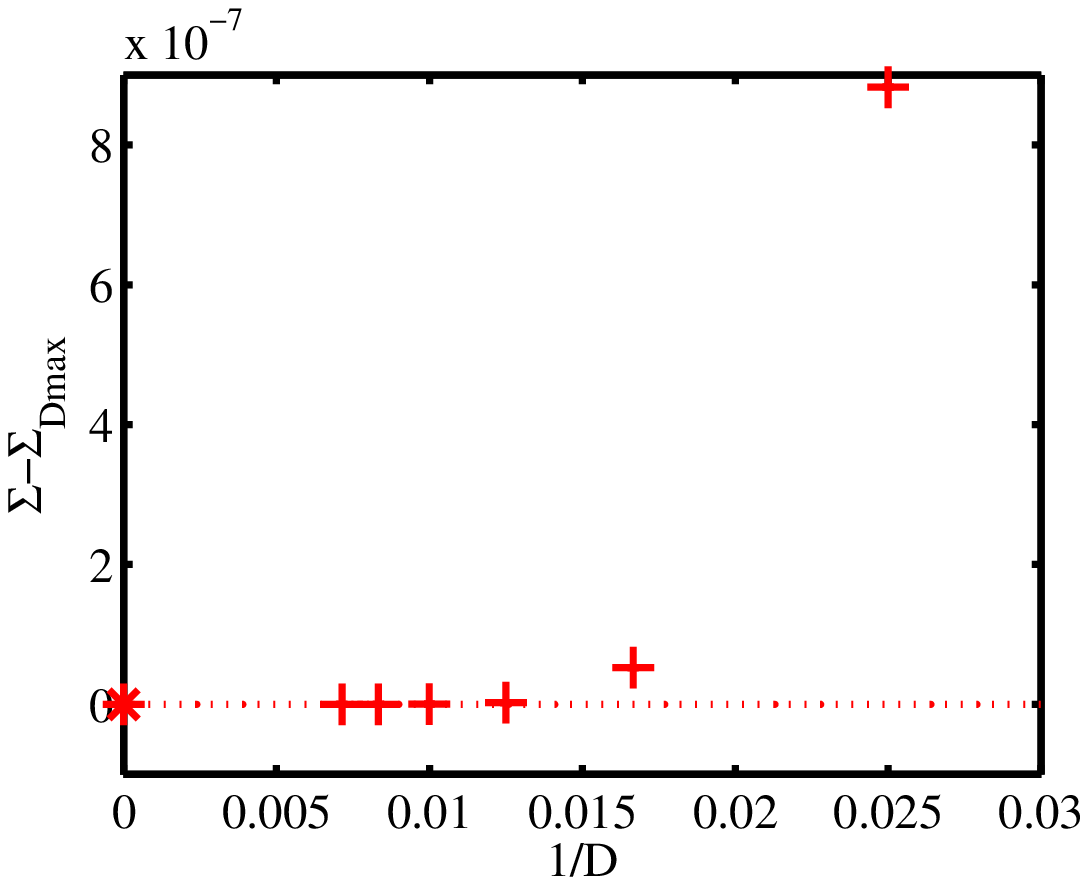}
}
\end{minipage}
\caption{Convergence of the condensate value with the bond dimension, $D$, for $x=100$ and $N=300$
for the massless (left) and a massive case (right). 
To better appreciate the convergence, we plot the difference between the observable at each bond dimension 
and the one obtained for the maximum $D_{\mathrm{max}}=140$ (respectively $\Sigma_{D_{\mathrm{max}}}=0.174663$
and $0.484859$). 
Dashed lines show the extrapolation in $1/D$ (linear from the largest values of $D$), with the final value displayed as a star.
The error is estimated from the difference between this value and the one for the largest computed $D$.
}
\label{fig:Dscaling}
\end{figure}

\begin{figure}[floatfix]
\begin{minipage}[b]{.4\columnwidth}
\subfigure[$m/g=0$]{
\includegraphics[height=.9\columnwidth]{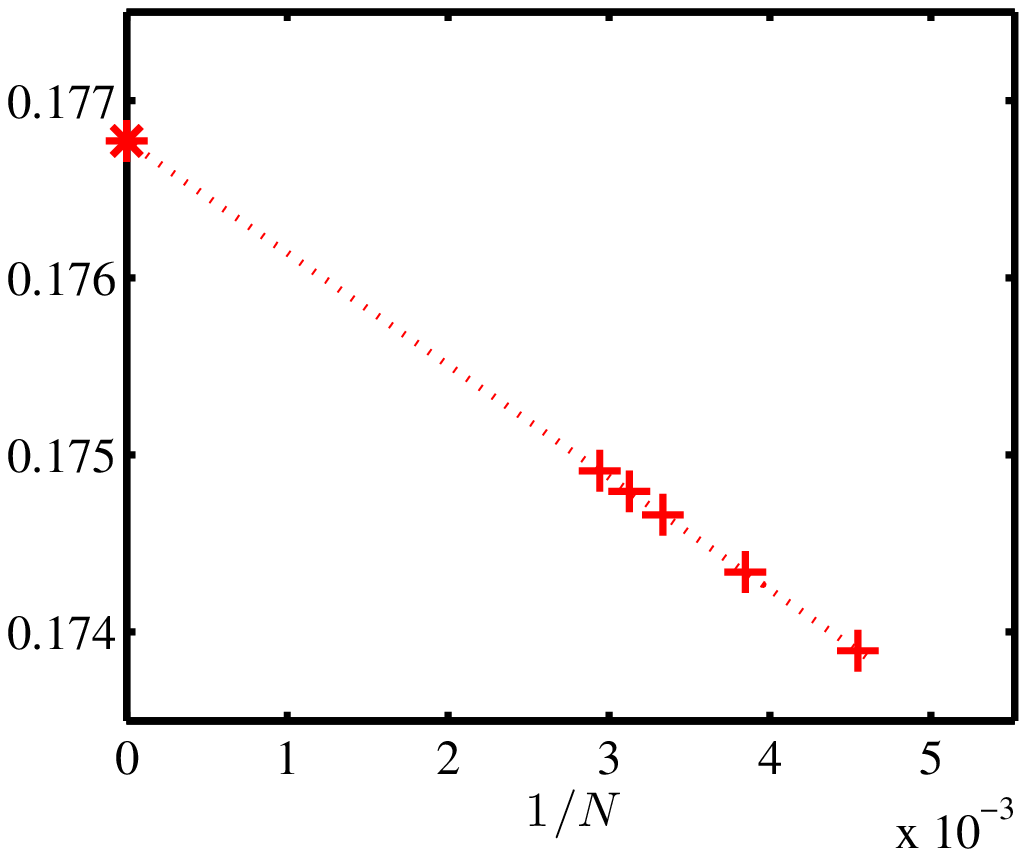}
}
\end{minipage}
\hspace{.1\columnwidth}
\begin{minipage}[b]{.4\columnwidth}
\subfigure[$m/g=0.25$]{
\includegraphics[height=.9\columnwidth]{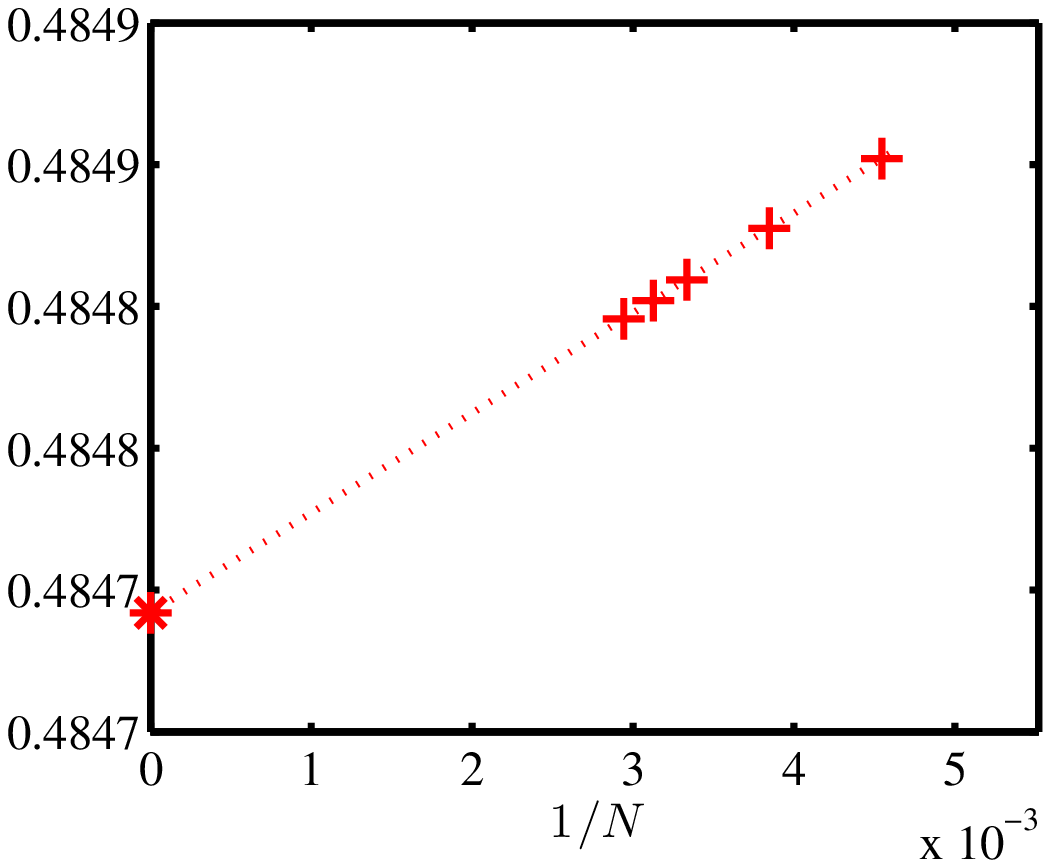}
}
\end{minipage}
\caption{Finite volume extrapolation of the (substracted) condensate for $x=100$
in the case of massless fermions (left) and $m/g=0.25$ (right).
We fit the results to a linear function in $1/N$, shown by a dashed line.
The extrapolated value is indicated by a star.
}
\label{fig:Lscaling}
\end{figure}

\begin{figure}[floatfix]
\begin{minipage}[b]{.4\columnwidth}
\subfigure[$m/g=0$]{  
\includegraphics[height=.9\columnwidth]{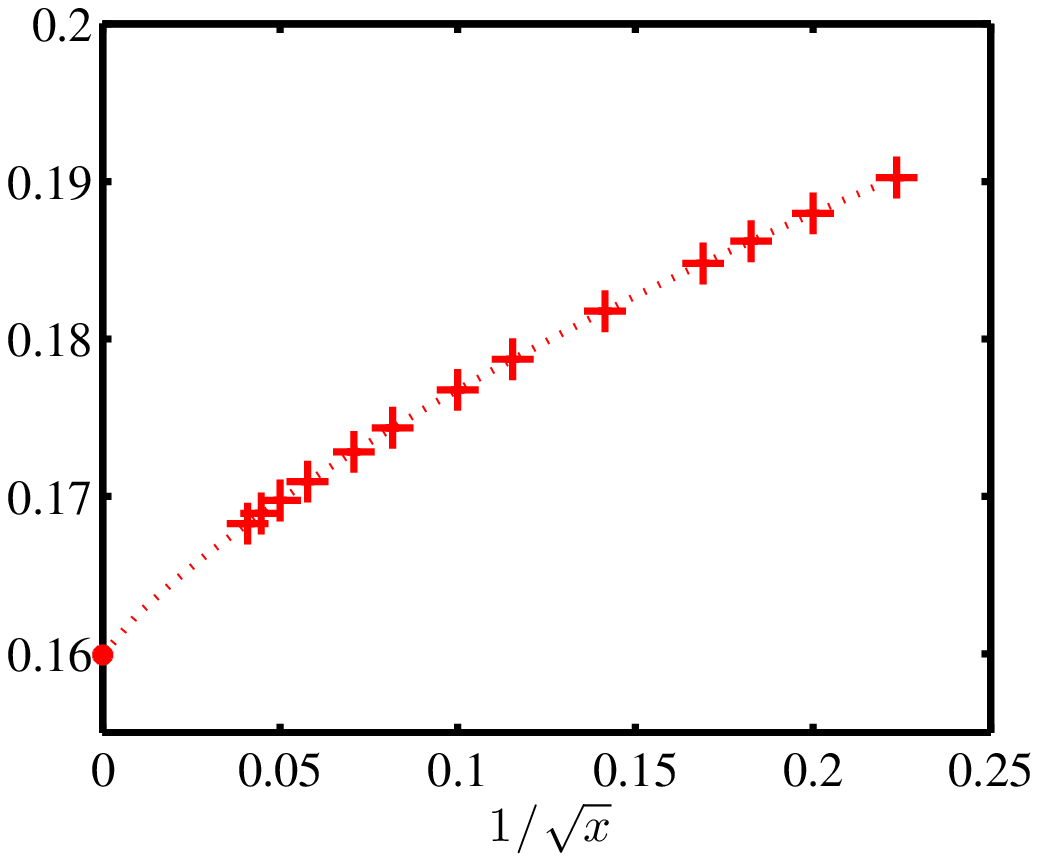}
}
\end{minipage}
\hspace{.1\columnwidth}
\begin{minipage}[b]{.4\columnwidth}
\subfigure[$m/g=0.25$]{ 
\includegraphics[height=.9\columnwidth]{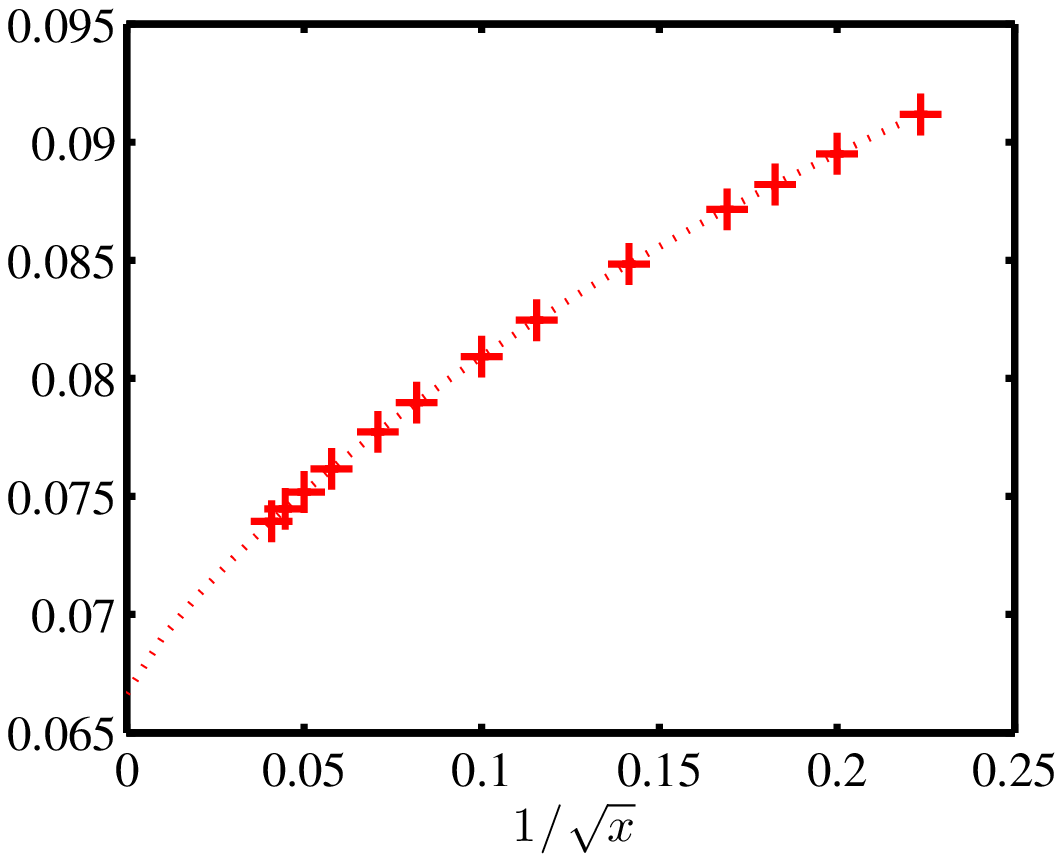}
}
\end{minipage}
\caption{
Substracted condensate as a function of $1/\sqrt{x}$ for fermion masses $m/g=0$ and $m/g=0.25$.
The error of each point (in most cases smaller than the markers) reflects the uncertainties of the linear
finite size extrapolations.
In the massless case, the exact result $\Sigma_0=e^{\gamma}/(2 \pi^{3/2})\approx 0.159929$
is indicated on the vertical axis.
The dashed lines corresponds to the fit of the whole computed range, $x\in[20,\,600]$.
}
\label{fig:substrCond}
\end{figure}

Here we repeat the analysis for the expectation value of $\hat{\Sigma}$ in the ground state.
Our results show very good convergence in the bond dimension, $D$, 
and we can reliably estimate the observable for a given set of parameters $(m/g,x)$ and a given system size, $N$,
from the linear extrapolation of the largest computed bond dimensions, as illustrated in Fig.~\ref{fig:Dscaling}.
From these values, we perform a finite size extrapolation using 
a linear function $f(N)=A+\frac{B}{N}$, which perfectly describes  our observations
(Fig.~\ref{fig:Lscaling}).
The error of the thermodynamic limit is estimated from the width of the $68\%$ confidence interval for the fitted parameter $A$.

From this thermodynamic values, for each given $(m/g,\,x)$ we subtract the exact free condensate in the bulk,
(\ref{eq:freecondbulk}).
Finally, the continuum limit extrapolation can be attained from all computed $x$ values.
According to the discussion in the previous section, divergent $\log a$ terms are not present in the
substracted condensate, but there might be residual $a \log a $ corrections introduced by the interaction, 
and which cannot be substracted. We thus use for the fit the form
\begin{equation}
f(x)=A+F\frac{\log x }{\sqrt{x}}+B\frac{1}{\sqrt{x}}+C\frac{1}{x},
\label{eq:fitCondX}
\end{equation}
which, as seen in Fig.~\ref{fig:substrCond}, describes our computed data extremely well.
Finally, we obtain for the condensate the following values.

\begin{center}
\begin{tabular}{|c|c|c|}
\hline
&
\multicolumn{2}{|c|}{Substracted condensate} \\
\hline
$m/g$  & MPS with OBC & exact \\
\hline
0 & 0.159930(8) & 0.159929 \\
\hline
0.125 & 0.092023(4) & - \\
\hline
0.25 &  0.066660(11) & - \\
\hline
0.5 & 0.042383(22) & - \\
\hline

\end{tabular}
\end{center}

\section{Discussion}

We have extended our previous study of the Schwinger model using MPS to the computation of the chiral
condensate.
Determining this quantity in the continuum is a more challenging task than finding energy levels, 
as evidenced by the few numerical estimations found in the literature.

We obtain a remarkably accurate values of the condensate in the massless case, where we can compare our
estimate to the exact result. 
For the massive cases, there are not many previous calculations in the literature, or the ones available correspond
to different masses and thus do not allow for a direct comparison to our results. 
In the case of $m/g=0.125$, we are in good agreement with the approximate value $0.0929$,
from \cite{hosotani98}.

Our results show the feasibility of the MPS ansatz to describe the physical states relevant for a lattice gauge theory problem.
The technique allows us to reach precisions that suffice for accurate finite size and continuum extrapolations, and to 
identify the asymptotic approach to the continuum.
These results pave the way to further applications of MPS or more general tensor network techniques
to more challenging problems in the context of LGT.

\subsubsection*{Acknowledgements}

This work was supported in part by the DFG Sonderforschungsbereich/Transregio SFB/TR9
and  by the EU through SIQS grant (FP7 600645).
K.C. was supported by Foundation for Polish Science fellowship ``Kolumb''.
H.S. was supported by the Japan Society for the Promotion of Science for Young Scientists.

\bibliography{lat13-mps}


\end{document}